\documentclass[useAMS]{mn2e}
\usepackage{psfig}

\usepackage{times}
\def\eg{\mbox{$E_\gamma$}}
\def\ega{\mbox{$E_{\rm \gamma,a}$}}
\def\ep{\mbox{$E_{\rm p}$}}
\def\eiso{\mbox{$E_{\rm \gamma,iso}$}}
\def\tjet{\mbox{$t_{\rm jet}$}}
\def\ta{\mbox{$T_{\rm a}$}}
\def\tap{\mbox{$T^\prime_{\rm a}$}}
\def\epeg{\mbox{$E_{\rm p}-E_\gamma$}}
\def\epega{\mbox{$E_{\rm p}-E_{\rm \gamma,a}$}}
\def\epeiso{\mbox{$E_{\rm p}-E_{\rm \gamma,iso}$}}
\def\sw{{\it Swift}}

\title[The role of break--times as jet angle indicators] 
{The role of afterglow break--times as GRB jet angle indicators}

\author[Nava et al.]
{L. Nava$^{1,2}$\thanks{E--mail: lara.nava@brera.inaf.it},
G. Ghisellini$^1$, G. Ghirlanda$^1$, J.I. Cabrera$^3$, C. Firmani$^{1,3}$ and 
\newauthor 
V. Avila--Reese$^3$\\
$^{1}$Osservatorio Astronomico di Brera, via E.Bianchi 46, I-23807
Merate, Italy\\
$^{2}$Universit\'a degli Studi dell'Insubria, Dipartimento di Fisica e Matematica, via Valleggio 11, I-22100 Como, Italy\\
$^{3}$Instituto de Astronom\'ia, Universidad Nacional Aut\'onoma de M\'exico, A.P. 70-264, 04510, M\'exico, D.F., M\'exico}

\begin{document}

\date{Accepted 2007 February 27. Received 2007 February 23; in original form
  2007 January 24}


\maketitle

\label{firstpage}

\begin{abstract}
  The early X--ray light curve of Gamma Ray Bursts (GRBs) is complex, and
  shows a typical steep--flat--steep behaviour.  The time \ta\ at which the
  flat (plateau) part ends may bear some important physical information,
  especially if it plays the same role of the so called jet break time \tjet.
  To this aim, stimulated by the recent analysis of Willingale et al., we have
  assembled a sample of GRBs of known redshifts, spectral parameters of the
  prompt emission, and \ta.  By using \ta\ as a jet angle indicator, and then
  estimating the collimation corrected prompt energetics, we find a
  correlation between the latter quantity and the peak energy of the prompt
  emission.  However, this correlation has a large dispersion, similar to the
  dispersion of the Amati correlation and it is not parallel to the Ghirlanda
  correlation.  Furthermore, we show that the correlation itself results
  mainly from the dependence of the jet opening angle on the isotropic prompt
  energy, with the time \ta\ playing no role, contrary to what we find for the
  jet break time \tjet.  We also find that for the bursts in our sample \ta\ 
  weakly correlates with \eiso\ of the prompt emission, but that this
  correlation disappears when considering all bursts of known redshift and
  \ta.  There is no correlation between \ta\ and the isotropic energy of the
  plateau phase.
\end{abstract}


\section{Introduction}

Before \sw, the afterglow observations were typically starting only several
hours after the burst, when the flux typically showed a single power law
decay.  The detection of GRB afterglows provided strong confirmation for the
fireball shock model (Rees \& Meszaros 1992), which explains it in terms of
forward shock emission running into the circumburst medium.  In many of the
well sampled afterglows, the steepening in the light curve at late times
($\sim10^5$ s) has been attributed to a narrow conical jet of semiaperture
opening angle $\theta_{\rm j}$ whose edges become visible as it decelerates
and widens (e.g. Rhoads et al. 1997).  The deceleration is due to the
expansion of the GRB outflow into the interstellar medium (ISM) whose
properties are still poorly understood.  By assuming a uniform density ISM the
widely dispersed isotropic energy \eiso\ of GRBs considerably clusters if
corrected for the collimation factor $f=(1-\cos\theta_{\rm j})$ (Frail et al.
2001).  Later, Ghirlanda, Ghisellini \& Lazzati (2004; hereafter GGL04)
discovered that the collimation corrected GRB energies
$E_{\gamma}=E_{\rm\gamma, iso} f$ are tightly correlated with the rest frame
peak energy \ep\ of the $\nu F_\nu$ prompt spectrum (so called ``Ghirlanda''
correlation).  Nava et al. (2006; hereafter N06) discovered that, in the case
of a wind--like circumburst density profile (which should be more likely if
the progenitor is a massive star), this correlation still exists and becomes
linear.  Liang \& Zhang (2005) discovered the phenomenological counterpart of
these correlations showing that a tight correlation exists between the \eiso,
\ep\ and \tjet\ (observed in the optical light curves) computed in the source
frame.  N06, indeed, demonstrated that this phenomenological correlation is
consistent with the \ep--\eg\ correlations computed either in the case of a
homogeneous and wind medium (HM and WM, respectively).

The jetted--fireball model predicts that when the bulk Lorentz factor
$\Gamma\sim 1/\theta_{\rm j}$, due to the deceleration of the expanding
fireball, an achromatic break appears in the afterglow light curve. Such
achromaticity is what distinguish a jet break (i.e. due to the geometry of the
event) from a spectral break due to the time--evolution of the
characteristic frequencies of the spectrum. Most of the jet breaks of the
pre--\sw\ GRBs were, indeed, achromatic, but in a narrow frequency band,
from the near--IR to the optical.
Moreover, the
typical jet breaks were observed between several hours to few days after the
GRB trigger: in order to identify this transition in the optical light curves
a systematic follow up of the optical afterglow, up to several days after the
jet break, was required. 

It was highly expected that the multi--wavelength \sw\ satellite 
(Gehrels et al. 2004) would clearly detect achromatic X--ray to optical breaks.  
Instead, \sw\ discovered
that the X--ray light curve of GRBs is much more complex than thought (e.g.
Burrows et al. 2005). 
The ``canonical'' X--ray light curve disclosed by the \sw\ 
observations is composed by an initial steep decay phase (commonly interpreted
as the prompt emission tail due to off axis radiation of a switching--off
fireball) followed by a shallow decay phase. 
At the end of this second phase
there is a break that marks the beginning of the ``normal'' decay phase,
observed also before \sw\ (i.e. when the X--ray follow up -- mostly with
{\it  Beppo}SAX -- started, several hours after the trigger).  
In addition,
\sw\ found that the X--ray afterglow is characterised by several early and
late times flares superposed to the typical power law flux decay (Nousek et al.
2006, O'Brien et al. 2006).  In several \sw\ bursts the optical light curve
does not track the X--ray one while in some cases flares and multiple breaks
have been observed in the optical too.  
Due to the richness and complexity of the light curves,
the identification of a break as a jet break requires care
(Ghirlanda et al. 2007 - G07 hereafter).

Some interpretations of the complex multi--break X--ray light curve of \sw\ 
GRBs have been proposed: while it is commonly accepted that the early steep
decay can be interpreted as due to the curvature effect of a switching--off
fireball (see e.g. Kumar \& Panaitescu 2000), 
the origin of the intermediate shallow decay is still 
controversial.
It has been proposed (Ghisellini et al. 2007) that flat X--ray could be ``late
prompt'' emission due to a late central engine activity producing shells with
a time--decaying bulk Lorentz factor $\Gamma$. 
In this scenario the flat--to--steep transition occurs when
$\Gamma$ of these late shells becomes $\sim 1/\theta_{\rm j}$.

Recently, Willingale et al. (2007; hereafter W07) interpreted the
steep--flat--steep X--ray light curves as due to the superposition of two
separate and independent components: the prompt emission (interpreted as
internal dissipation) produces the first steep decay while the afterglow
emission (external dissipation) is responsible for the shallow and the final
steep components. 
The modelling, introduced by W07, identifies an early break
(typically $\sim 10^3$--$10^4$ s) in the X--ray light curve, called \ta, that
marks the shallow--to--steep transition. W07 show that if \ta\ is used (with
the same formalism adopted for $t_{\rm jet}$ in the HM case) to compute an
angle $\theta_{\rm j, a}$, the collimation corrected energy 
$E_{\rm \gamma, a}=E_{\rm \gamma, iso}(1-\cos\theta_{\rm j, a})$ is 
correlated with \ep.  
W07 argue that this correlation has the same slope of the 
\epeg\ (Ghirlanda) correlation (found with the jet break time \tjet).
The two correlations are simply displaced by a factor $\sim 26$ in $E_\gamma$.
This implies that the two break times \ta~and \tjet\ are proportional
(i.e. $\tjet\sim 90 \ta$). 
In the sample of W07 only GRB 050820A has both \ta\ and \tjet\ measured,
and indeed in this case \tjet$\sim 100$\ta.
W07 conclude that if there is a link between \ta\ and \tjet, it is 
doubtful that both times are actually jet break times.  
However, W07 stress that it is possible that the end of the plateau
phase (corresponding to \ta) depends on the total energy of the outflow.
They also point out that it would be interesting to study
the phenomenological counterpart of the \epega\ 
correlation similarly to what has been done with the Liang--Zhang correlation
with respect to the Ghirlanda one.

The scope of this paper is to investigate the role of the break time \ta\ in
defining the \epega\ correlation and if a Liang--Zhang like correlation exists
between the quantities \ep, \eiso\ and \tap\ (where \ep\ and \tap\ are
expressed in the source rest frame).  To this
aim, we enlarge the sample used by W07 from 14 to 23
GRBs and derive the collimated corrected energy \ega\ using
\ta\ as a jet break and we reproduce the correlation between \ep\ and \ega\ and
test, through the scatter of these correlations, the nature of \ta.

We adopt a standard cosmology with $\Omega_{\rm m}=0.3$, 
$\Omega_\Lambda=h=0.7$.


\begin{table*} 
\begin{tabular}{llllllcl} 
\hline   
GRB     &$z$   &$E_{\rm p}$  &$E_{\rm \gamma, iso, 52}$  &Instr$^a$ &Ref$^b$ &$\log T_a^c$    &Ref$^d$\\
        &      &keV          & erg   &          &        &s               &   \\
\hline
050318  &1.44  &115$\pm$27   &2.03$\pm$0.34     &SWI     & 1     &2.01 [1.37--2.68]   &W07\\
050401  &2.90  &501$\pm$117  &41$\pm$8          &KON     & 2$^e$ &3.87 [3.69--4.11]   &W07\\
050416A &0.653 &28.6$\pm$8.3 &0.083$\pm$0.029   &SWI     & 3     &3.19 [2.69--3.48]   &W07\\
050525A &0.606 &127$\pm$5.5  &2.96$\pm$0.64     &SWI     & 4     &2.92 [1.69--3.06]   &W07\\
050603  &2.821 &1333$\pm$107 &59.8$\pm$4.0      &KON     & 5     &4.83 [3.69--5.13]   &W07\\
050820A &2.612 &1325$\pm$277 &97.8$\pm$8.0      &KON     & 6     &3.96 [3.84--4.04]   &W07\\
050922C &2.198 &417$\pm$118  &4.61$\pm$0.87     &HET     & 7     &2.58 [2.48--2.69]   &W07\\
        &...   &...          &...               &...     &...    &3.98 [3.68--4.16]   &G07\\
051109A &2.346 &539$\pm$381  &7.58$\pm$0.94     &KON     & 8     &3.93 [3.70--4.08]   &W07\\
060115  &3.53  &281$\pm$82   &8.0$\pm$1.5       &SWI     & 9     &3.86 [2.86--5.47]   &W07\\
060124  &2.297 &636$\pm$162  &43$\pm$4.0        &KON+SWI & 10    &4.60 [4.47--4.69]   &W07\\
         &...   &...          &...               &...     &...    &3.70 [3.40--3.88]   &this paper, see Fig. \ref{ta}\\
060206  &4.048 &381$\pm$98   &4.75$\pm$0.76     &SWI     & 11    &3.86 [3.68--4.00]   &W07\\
060418  &1.489 &572$\pm$114  &12.8$\pm$1.1      &KON     & 12    &3.44 [3.26--3.58]   &W07\\
060526  &3.21  &105$\pm$21   &2.58$\pm$0.26     &SWI     & 13    &3.84 [3.51--4.31]   &W07\\
060707  &3.43  &292$\pm$71   &6.83$\pm$1.49     &SWI     & 14    &3.58 [2.87--4.05]   &W07\\
060927  &5.6   &473$\pm$116  &9.69$\pm$1.62     &SWI     & 15    &3.60 [3.30--3.78]   &this paper, see Fig. \ref{ta}\\
061121  &1.314 &1289$\pm$153 &26.1$\pm$3.0      &KON/RHE & 16-17 &3.90 [3.60--4.08]   &this paper, see Fig. \ref{ta}\\  
\hline
050126  &1.29  &202$\pm$49   &1.10$\pm$0.30     &SWI     & this paper &2.34 [1.34--5.64]   &W07\\
050505  &4.27  &622$\pm$211  &19.5$\pm$3.1      &SWI     & this paper &4.39 [4.15--4.87]   &W07\\
050908  &3.344 &191$\pm$40   &2.05$\pm$0.34     &SWI     & this paper &3.31 [2.03--4.17]   &W07\\ 
060223A &4.41  &341$\pm$69   &4.54$\pm$0.72     &SWI     & this paper &2.73 [2.37--3.05]   &W07\\ 
060522  &5.11  &415$\pm$78   &8.02$\pm$1.65     &SWI     & this paper &2.86 [2.61--3.32]   &W07\\ 
060607A &3.082 &535$\pm$164  &12.2$\pm$1.8      &SWI     & this paper &4.75 [4.73--4.78]   &W07\\ 
060908  &2.43  &487$\pm$116  &9.17$\pm$1.57     &SWI     & this paper &3.10 [2.80--3.28]   &this paper, see Fig. \ref{ta}\\
\hline

\end{tabular} 
\caption{GRB sample used to study the \epega\ correlation. The peak energy 
  \ep\ is in the source rest frame. The time \ta\ is in the observer frame. 
  \eiso\ is given in units of 10$^{52}$ erg. 
  $^a$The $+$ sign indicates that the spectral parameters are derived  from the
  joint spectral fit of two different instruments. A sign / indicates that the
  reported values of \ep and \eiso\ are an average of the 
  values found by two different instruments. 
  The instruments are: SWI=\sw/BAT, KON=Konus/Wind,
  HET=HETE/Fregate, RHE=RHESSI. 
  $^b$References for the spectral parameters:  (1)
  Perri et al., 2005; (2) Golenetskii et al., 2005a; (3) Sato et al., 2006; (4)
  Blustin et al., 2006; (5) Golenetskii et al., 2005b; (6) Cenko et al., 2006; (7)
  Crew et al., 2005; (8) Golenetskii et al., 2005c; (9) Barbier et al., 2006; (10)
  Romano et al., 2006; (11) Palmer et al., 2006; (12) Golenetskii et al., 2006a;
  (13) Schaefer 2007; (14) Stamatikos et al., 2006a; (15) Stamatikos et al., 2006b; 
  (16) Golenetskii et al., 2006b; (17) Bellm et al., 2006. 
  $^c$Values in square brackets are the lower and the upper 90\% confidence
  limits. 
  $^d$References for \ta: Willingale et al. 2007 (W07); 
  Ghirlanda et al. 2007 (G07). 
  $^e$The values of \ep\ 
  and \eiso\ derive from the average of the two spectral fit presented
  in the corresponding reference. For GRB~050922C and 060124 we report the value 
  of \ta\ given in W07 (first line) and the value obtained in G07 and 
  this paper (second line), respectively.}  
\label{tab}
\end{table*}

\section{The sample}
\label{sample}

In order to build the \epega\ correlation we need the redshift $z$, the peak
energy of the prompt emission spectrum \ep\ and, from the X--ray
light curve, the time \ta.  We found, up to November 2006, 16 objects with
all these quantities published in the literature.  We added 7 more GRBs
(050126 050505 050908 060223A 060522 060607A 060908) for which we analysed the
\sw--BAT (15--150 keV) spectrum and could satisfactorily constrain the spectral
peak energy $E_{\rm p}$ fitting the spectrum with a cut-off power law model.
In these cases we also tested our results (by the $F$--test) finding that the
fit with a cut--off power law is better than a single power law.  The details
of the fitting, together with the results for other \sw\ bursts, will be
given elsewhere (Cabrera et al. 2007 in preparation).

The sample of W07 which is used to build the \epega\ correlation contains 14
long GRBs.  However, it is hard to identify those bursts present in both
samples because W07 do not give the names of those bursts they use to compute
the \epega\ correlation.

\begin{figure*}
\begin{tabular}{cc}
  \hskip -0.3 cm
  \psfig{figure=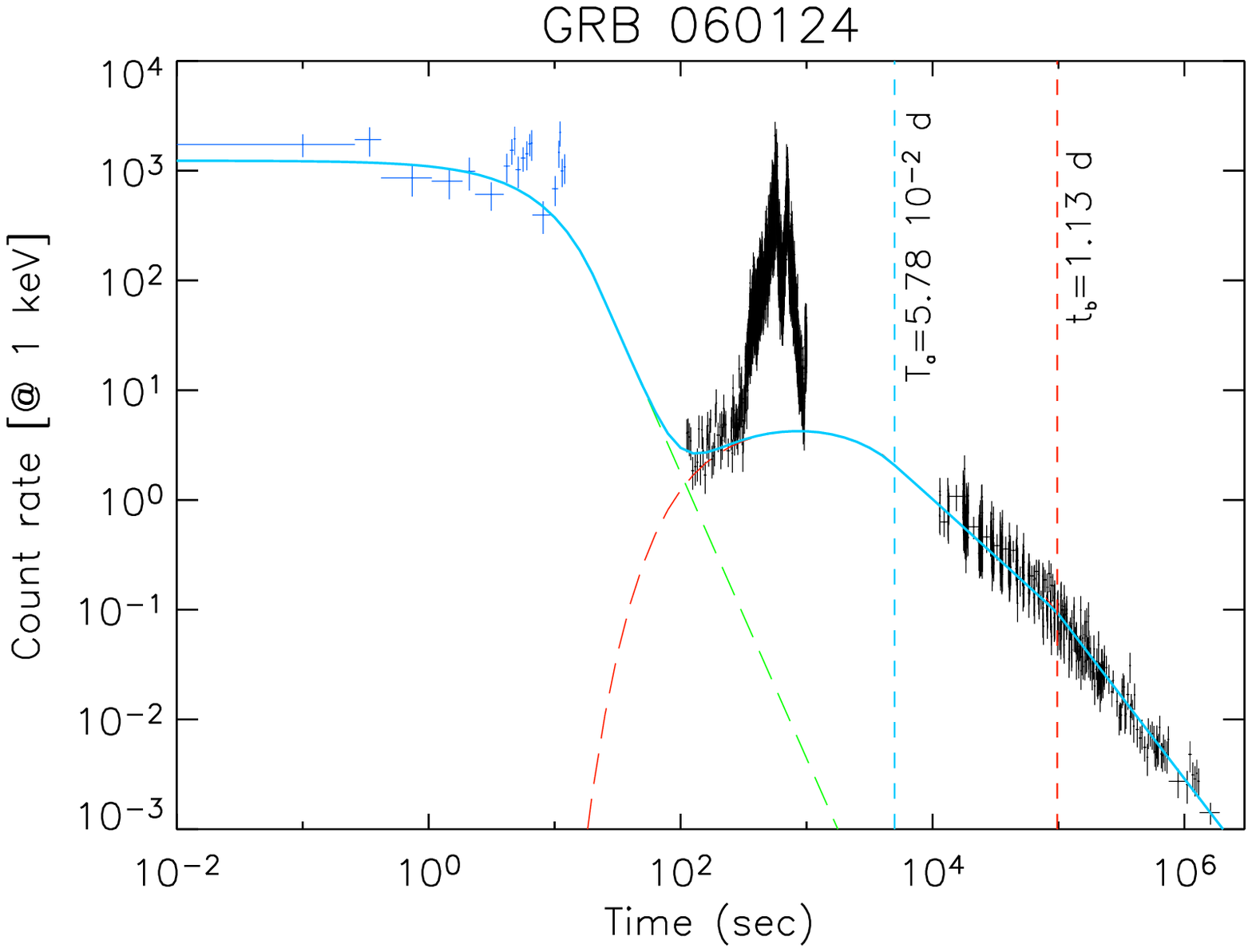,width=8.0cm,height=6.cm}
  \psfig{figure=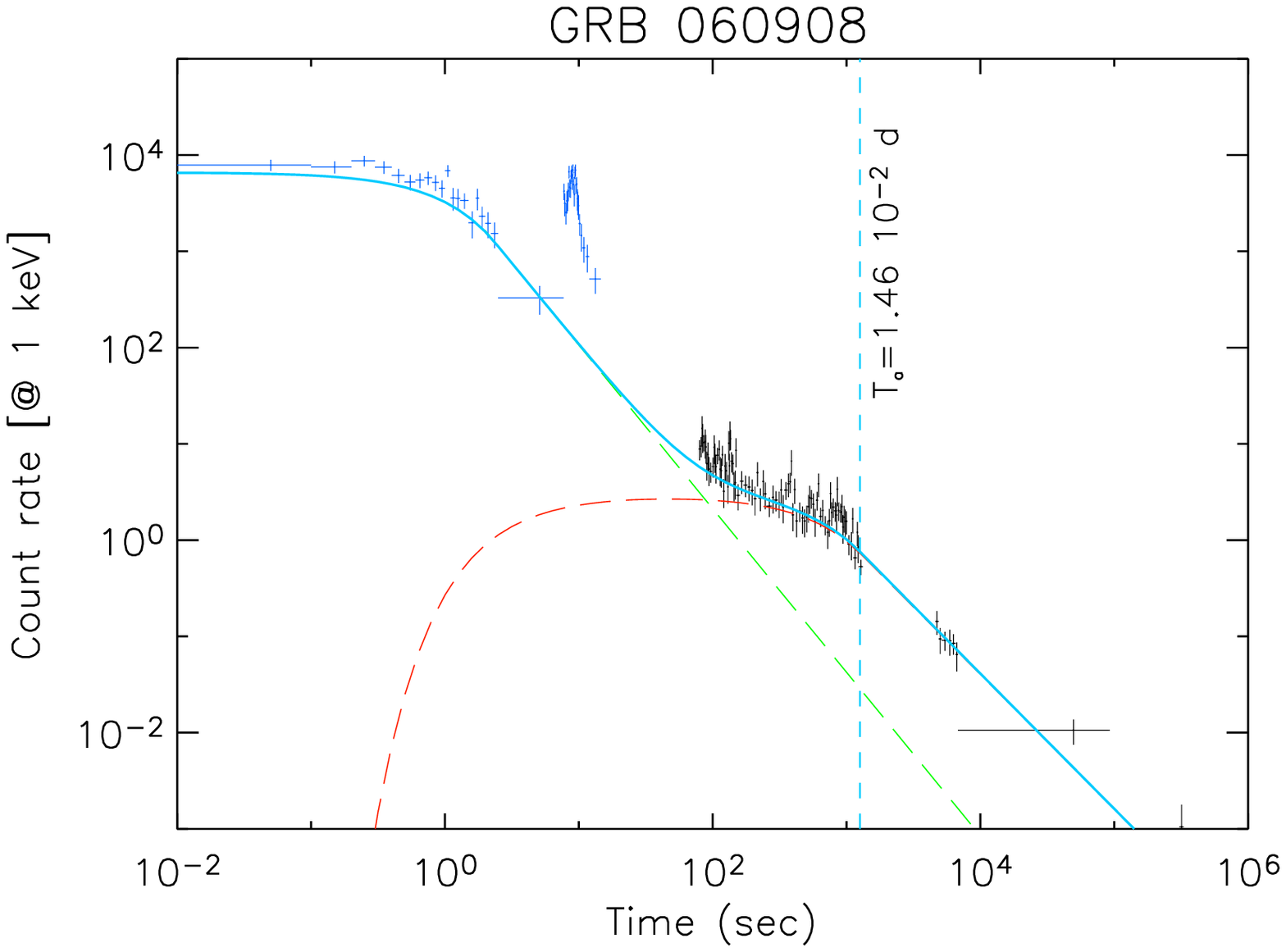,width=8.0cm,height=6.cm}\\
  \hskip -0.3 cm
  \psfig{figure=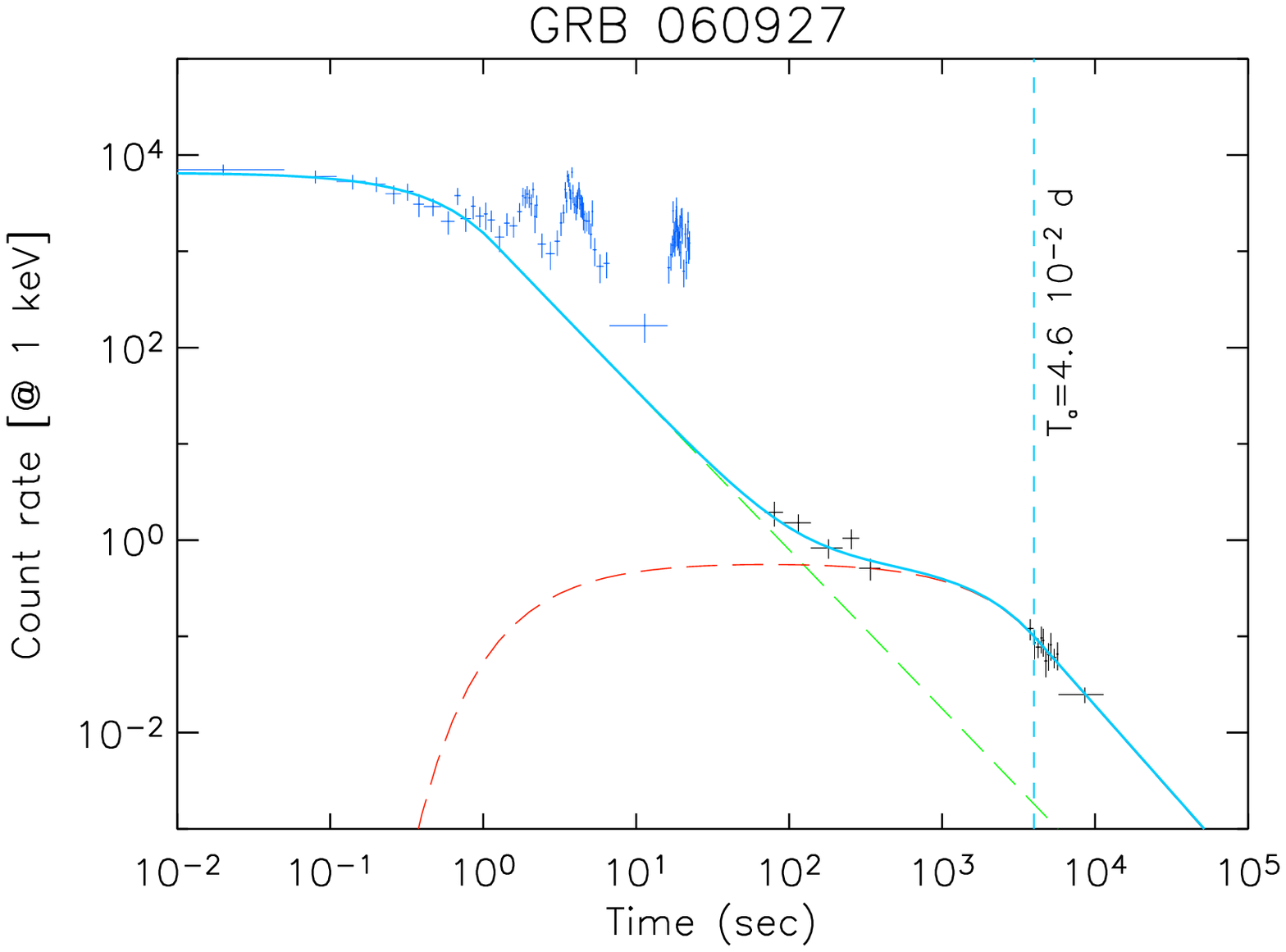,width=8.0cm,height=6.cm}
  \psfig{figure=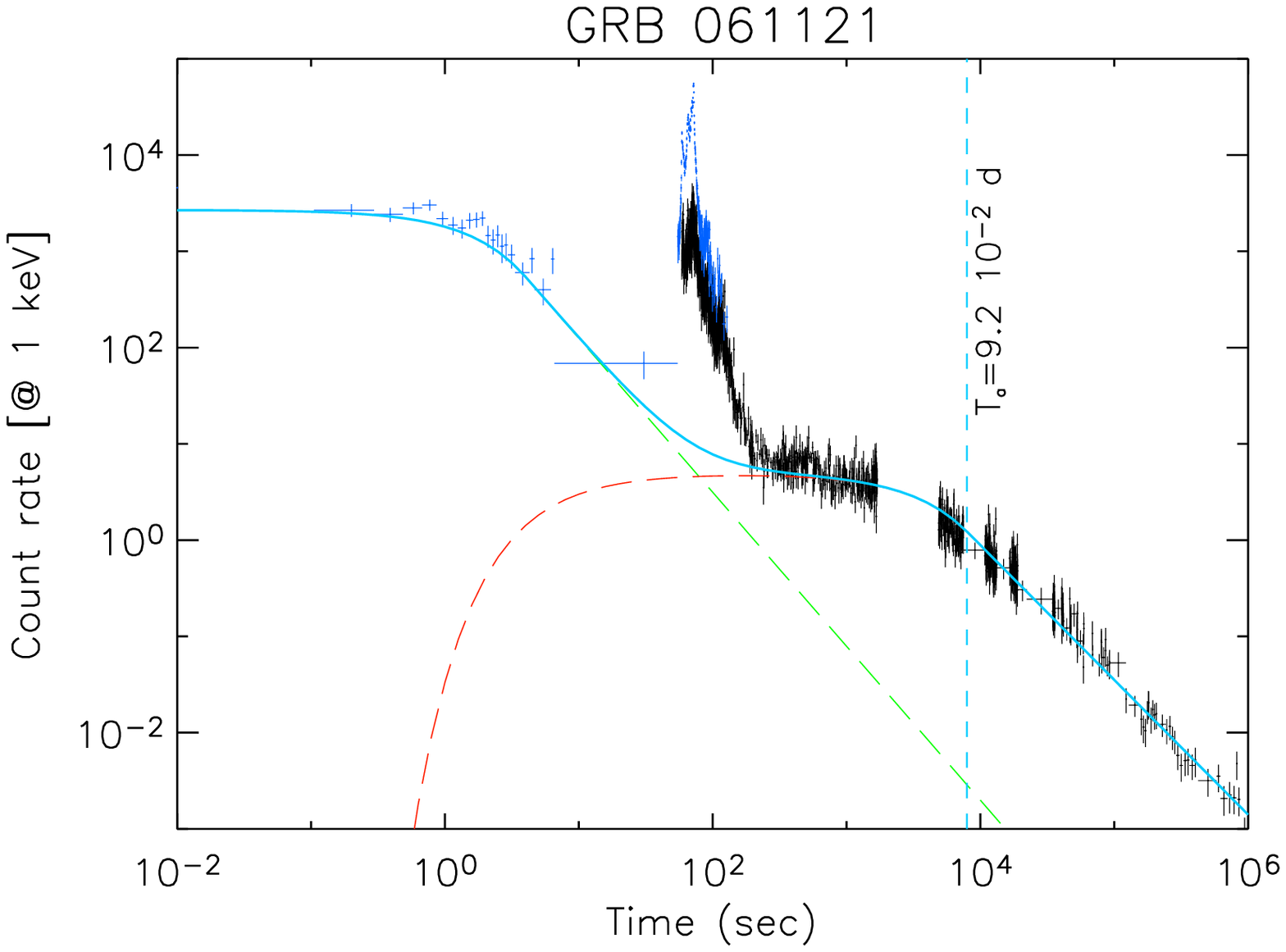,width=8.0cm,height=6.cm}
\end{tabular}
\caption{
  The early BAT and late XRT light curves of GRB 060124, GRB 060908, GRB
  060927 and GRB 061121 modelled with the function proposed by W07. The two
  model components are shown by the long--dashed lines while their sum is
  shown by the solid line. The fits were performed excluding the flares of the
  light curves. The long dashed vertical lines show the \ta\ in each light
  curve (for GRB~060124 we also report the jet break time found in the optical
  light curve -- see text). }
\label{ta}
\end{figure*}

Our sample is composed by 23 GRBs whose properties are listed in Table
\ref{tab}.  
In most cases the value of \ta\ is taken from the
compilation of W07, except for two cases:
\begin{itemize}
\item for GRB 050922C we have found that the X--ray light curve does not
  clearly show the shallow part and, therefore, the value of \ta\ is hardly
  constrained. A still acceptable fit is obtained by setting \ta$\sim$10$^4$
  seconds (a factor 26 larger than what derived by W07 -- see G07);
\item for GRB 060124 the fit of the X--ray light curve should account for the
  late time achromatic break (Romano et al. 2006; Curran et al. 2006) which
  corresponds to the jet break time (G07). With the
  inclusion of this further late time break \ta$\sim 5000$ s (a factor 8
  smaller than what reported in W07).  The fit is shown in Fig. \ref{ta}.
\end{itemize}

Moreover, since the sample of W07 comprises GRBs up to August 2006, for GRB
060908, GRB 060927 and GRB 061121, we have estimated \ta\ following the
procedure outlined in W07, and using the publicly available X--ray
light--curves\footnote{ http://astro.berkeley.edu/$\sim$nat/swift/ by N.
  Butler, see also Butler \& Kocevski (2007)}.  
The corresponding light curves and fitting models are shown in Fig. \ref{ta}.

In the following sections we analyse the \epega\ correlation (found using \ta\
as an angle estimator) and compare it with the \epeg\ correlation (found using
\tjet\ as angle indicator). We also study the relevance of either \ta\ and
\tjet\ in ``collapsing" the scatter of the Amati correlation 
(Amati et al., 2002; Amati 2006), 
in the respective collimation corrected correlations.

For the \epega\ correlation we use the sample of 23 GRBs of the \sw--era
reported in Tab. 1.  For the comparison with the \epeg\ correlation we use the
most updated sample of 23 bursts for which \tjet\ has been firmly identified
from the optical light curve reported in G07\footnote{In the present analysis
we do not include the lower limits on \tjet\ discussed in G07.}.  
Within
this sample (with \tjet) there are 6 GRBs of the \sw--era for which we also
have an estimate of \ta. 
To reproduce the Amati correlation (which only
requires the knowledge of the redshift and of the spectral parameters) we have
used those GRBs that appear at least in one of the two samples. This
larger sample contains 40 GRBs.

\section{Analysis and results}
\label{aar}

First we study the Ghirlanda like correlation and the Liang \& Zhang like
correlation defined using \ta\ in place of \tjet.  
We use the formalism defined in N06 where, 
for a generic time break $t_{\rm break}$,
the jet opening angle is defined as{\footnote{We use the notation
$Q_x=10^x Q$, using cgs units.}:
\begin{eqnarray}
\theta  &=&  0.161 \left({ t_{\rm break} \over 1+z}\right)^{3/8}
\left({n \, \eta_{\gamma}\over E_{\rm \gamma,iso,52}}\right)^{1/8}
   \, \, \,\,\,\,\, {\rm [HM]} \nonumber \\
\theta  &=&  0.2016 \,
\left( {t_{\rm break} \over 1+z}\right)^{1/4}
\left( { \eta_\gamma\ A_* \over E_{\rm \gamma,iso,52}}\right)^{1/4}
   \, \, \, {\rm [WM]}
\label{ang}
\end{eqnarray}
where $n$ represents the circumburst density in the homogeneous case (HM) and
$A_{*}$ is the normalisation of the wind (WM) density profile ($n(r)\propto
A_{*}r^{-2}$ -- see N06 for more details).  Here $\eta_{\gamma}$ is the
radiative efficiency of the prompt phase assumed to be 0.2 for all bursts.

In both cases of Eq. \ref{ang}, it is necessary to know the isotropic
bolometric rest frame equivalent energy \eiso. 
This is
\begin{equation}
E_{\rm \gamma,iso}\, =\, {4\pi d^2_L S_{\gamma} k \over 1+z}
\label{eiso}
\end{equation}
where $d_L$ is the luminosity distance, $S_\gamma$ is the $\gamma$--ray
fluence in the observed energy band, $k$ is the bolometric correction factor
needed to find the energy emitted in a fixed energy range (here, 1--10$^4$
keV) in the rest frame of the source. For most pre--\sw\ 
bursts the spectrum is fitted with the Band function (Band et al. 1993)
composed by two smoothly joint power laws. 
On the other hand, several \sw\ 
bursts, due to the narrow energy range of the BAT instrument (15--150 keV),
have a spectrum best fitted with a cutoff--power law model. As discussed in
Firmani et al. (2006), in these cases the extrapolation of the
cutoff--power law model up to 10$^4$ keV may considerably underestimate the
energy, if the spectrum is, instead, a Band function. 
To account for this
limitation of the spectral fits in the \sw\ era, the value of \eiso\
reported in Tab. 1 is an average between the value derived with the
cutoff--power law model and that derived assuming a Band model with high energy
photon index $\beta=-2.3$ (see also G07).

\subsection{Correlation analysis}

For the 23 GRBs of Tab. 1 we computed the angle $\theta_{\rm j, a}$, where the
subscript ``a'' means that the angle has been estimated using \ta\
as a jet break time, and the collimation corrected energy $E_{\rm \gamma,a}$ 
in both the WM and HM case. 
The density in the homogeneous case is known only for some GRBs and
for the others we take the value $n=3\pm2.76$; the density in the wind case is
assumed equal for all GRBs and without error (see N06 and G07 for details).

We report also the correlation defined with the jet break time \tjet\ and
recently updated with the \sw\ bursts in G07. 
Here we use the sample of G07 (Tab. 1 in that paper) 
excluding bursts with only a lower limit on \tjet.  For
comparison we also recomputed the \epeiso\ (Amati)
correlation with the sample of 40 GRBs.

The fit is performed using the routine {\it fitexy} (Press et al. 1999) which
weights for the errors on the involved variables. 
All the fit results can be expressed in the synthetic form:
\begin{equation}
\left({E_{\rm p} \over 100\, {\rm keV}}\right) \, =\, 
(K\pm\sigma_{\rm K})\, 
\left({E_{\rm \gamma,a}\over E_{\rm bar}}\right)^{s\pm
  \sigma_{\rm s}}
\label{gen}
\end{equation}
where \ega\ must be replaced by \eg\ or \eiso\ according to the correlation
considered. Here $E_{\rm bar}$ represents the barycentre of the energy values
of the sample considered.

The three correlations are shown in Fig. \ref{trecorreleh} (homogeneous case)
and in Fig. \ref{trecorrelew} (wind case), and the fit results are listed in
Tab. \ref{tab2}. For every correlation we report also the reduced $\chi^2$ and
the 1$\sigma$ scatter of the data points around the best fit line, modelled
with a gaussian distribution.

\begin{figure*}
  \vskip -0.4 cm 
  \centerline{
  \psfig{figure=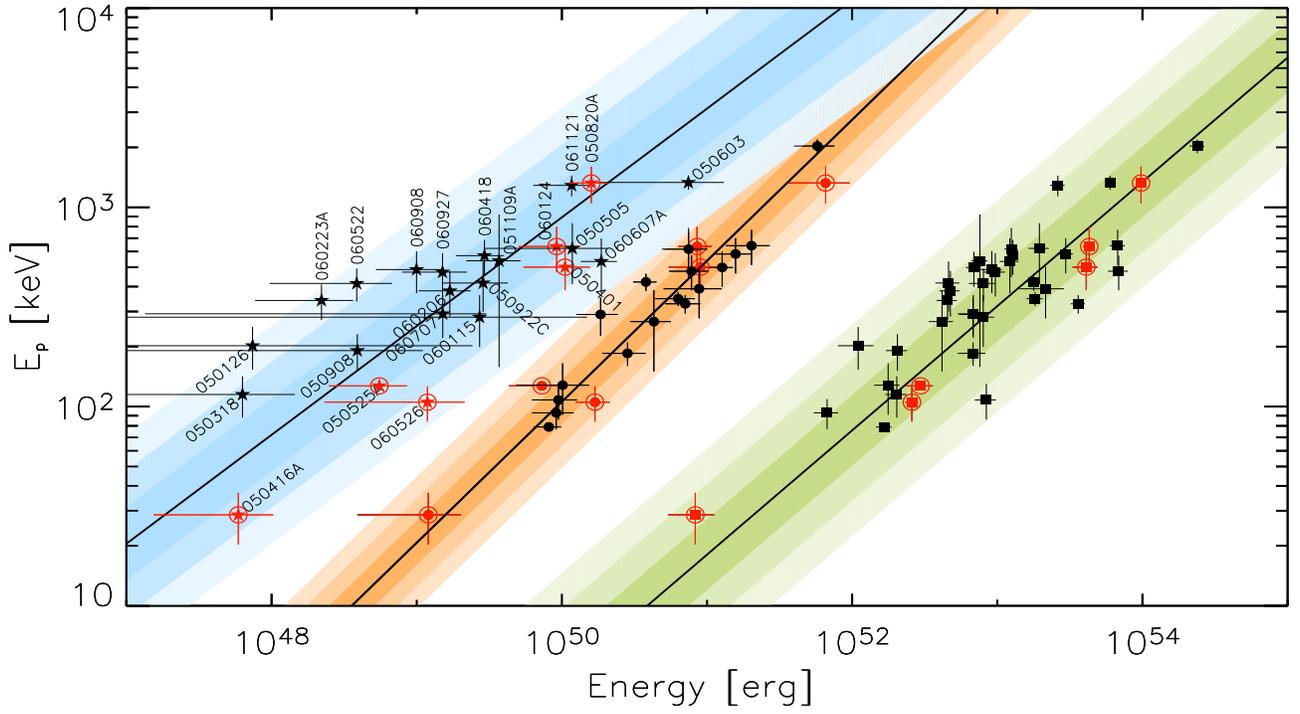,width=19.2cm,height=10.5cm}}
\caption{Homogeneous circumburst medium. From right to left 
  we show the \epeiso\ correlation (filled squares -- 40 GRBs), the
  \epeg\ correlation (filled circles -- 23 bursts from G07) and the \epega\ 
  correlation (filled stars -- 23 GRBs of Tab. 1). The shaded regions (for the
  three correlations) represent the 1, 2, 3$\sigma$ scatter of the data points
  (computed perpendicular) with respect to the best fitted correlation (solid
  lines).  Circled points (on the three correlations) represent  the 6 \sw\ 
  bursts for which both \ta\ and \tjet\ are known.}
\label{trecorreleh}
\end{figure*}
\begin{figure*}  
  \vskip -0.5 cm 
  \centerline{
  \psfig{figure=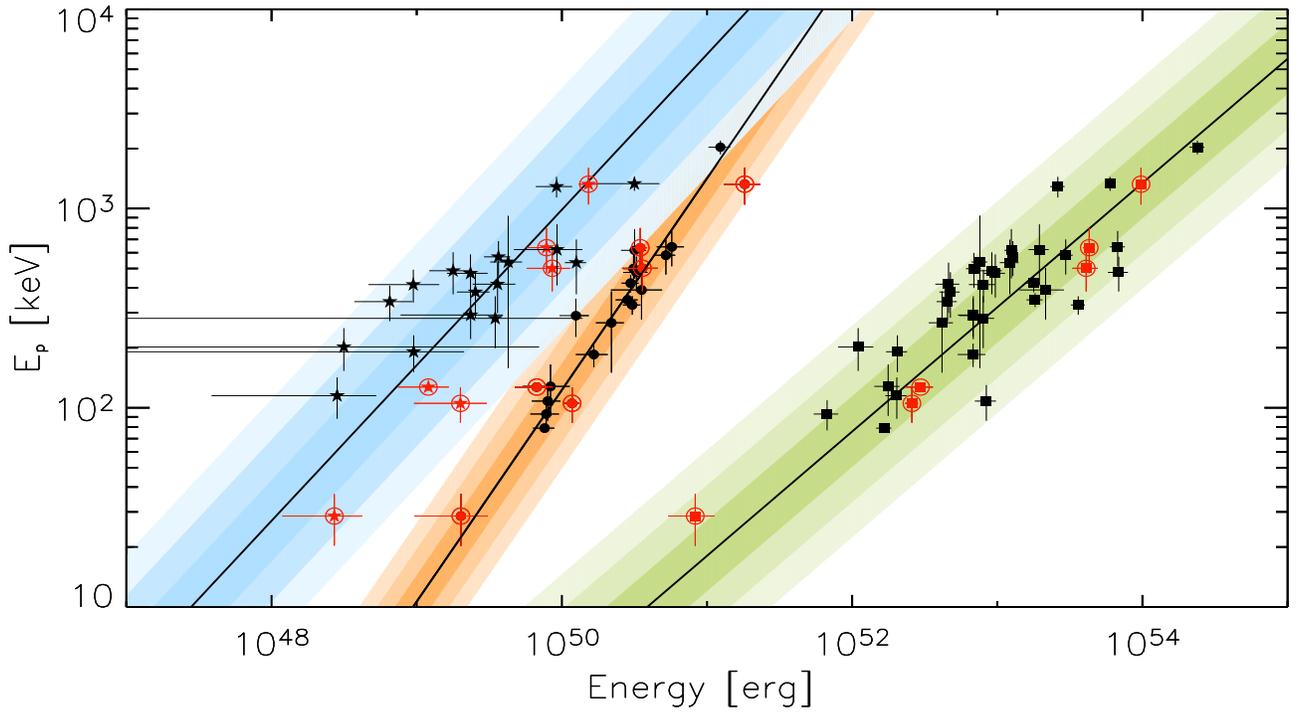,width=19.2cm,height=10.5cm}}
\caption{Wind circumburst medium. Symbols as in Fig.\ref{trecorreleh}}
\label{trecorrelew}
\end{figure*}

\begin{table*} 
\begin{tabular}{llccccccc} 
\hline   
Correlation & Density &N & $K$                 & $s$             & $E_{\rm bar}$     & $\chi^2_{\rm r}$   & Scatter \\
            & profile & & ($\sigma_{\rm K}$)& ($\sigma_{\rm s})$&     (erg)         &                    &         \\
\hline
\epega      & homog.  &23 & 4.71 (0.37)       & 0.55 (0.05)       & $3.11\times10^{49}$ & 1.95 (21 d.o.f.) &  0.16   \\
\epega      & wind    &23 & 4.95 (0.38)       & 0.78 (0.07)       & $4.14\times10^{49}$ & 1.95 (21 d.o.f.) &  0.16   \\
\epeg       & homog   &23 & 2.89 (0.16)       & 0.71 (0.04)       & $4.14\times10^{50}$ & 1.14 (21 d.o.f.) &  0.09   \\
\epeg       & wind    &23 & 3.21 (0.17)       & 1.06 (0.07)       & $2.47\times10^{50}$ & 0.91 (21 d.o.f.) &  0.08   \\
\epeiso     & ------  &40 & 3.77 (0.11)       & 0.62 (0.02)       & $1.31\times10^{53}$ & 5.70 (38 d.o.f.) &  0.16   \\ 
\epeiso     & ------  &23 & 4.83 (0.21)       & 0.65 (0.03)       & $1.22\times10^{53}$ & 2.89 (21 d.o.f.) &  0.15   \\ 
\hline 
\end{tabular} 
\caption{
Best fit results of the correlations involving the rest frame 
peak energy and the collimation corrected energy computed either using 
\ta\ as a jet break time and the real jet break time \tjet. 
We also report the Amati correlation for the considered 40 GRBs
and for the 23 defining also the \epega\ correlation.
The collimation corrected correlations are reported 
for both the homogeneous and wind medium scenario. $K$ and $s$ represent 
the normalisation and slope, respectively, of the best fit correlation 
(see Eq. \ref{gen}). The 1$\sigma$ scatter of the data points around the 
best fit correlation is also reported. }
\label{tab2}
\end{table*}

~ From the comparison of the results reported in Tab. 2 and in
Figs. \ref{trecorreleh} and \ref{trecorrelew} we note:
\begin{enumerate}
\item the correlation \epega\ derived using \ta\ as a jet break time is not
  parallel to the \epeg\ correlation (derived using \tjet). In particular,
  independently from the specific density profile considered, the \epega\ 
  correlation is much flatter than the \epeg\ correlation and in the
  homogeneous case it is even slightly flatter than the Amati correlation.
  By reconstructing the sample of W07 (through the quoted GCNs and the values
  given in that paper), and analyzing this smaller sample, we indeed
  obtain a \epega\ correlation parallel to the \epeg\ one, although
  not as tight.
  It is the inclusion of the additional GRBs that changes the slope of
  \epega\ correlation and further increases its scatter. 
  Also the 6 GRBs with both \ta\ and \tjet\ (encircled symbols)
  seem to define a  \epega\ parallel to the \epeg\ correlation, and
  this issue will be discussed in Sec. 3.2.
\item the fit of the \epega\ correlation (in terms of $\chi^2$) is
  significantly worse than that of the \epeg\ correlation, which, instead, is
  statistically acceptable;
\item the \epega\ correlation has a scatter which is a factor 2 larger than
  that of the \epeg\ correlation and is similar to that of the Amati
  correlation.
\end{enumerate}

Item (i) suggests that the relation between \tjet\ and \ta, if any, is not
trivial (i.e. \tjet\ is not simply proportional to \ta\ as suggested by W07).
The large scatter of the \epega\ correlation, which is comparable to that of
the \epeiso\ correlation, raises the question whether the new added variable
\ta\ can be treated as a jet break time or not. If not, we are still left with
the possibility that a completely phenomenological correlation between \ep\ 
and \eiso\ and \tap\ exists (where \tap\ is the time in the source rest
frame).  This possibility will be explored in the next section.

\subsection{The \ep--\eiso--\tap\ correlation}

In this section we study the \ep--\eiso--\tap\ correlation. First we fit the
multi--variable correlation with the 23 GRBs of Tab. 1 through a least--square
method without considering the errors on the three variables. The result is:
\begin{eqnarray}
E_{\rm \gamma,iso, 52} \, &=& (7.29\pm 0.89)\, 
\left(E_{\rm p} \over 357\, {\rm keV} \right)^{1.5\pm 0.1} \,  \times
\nonumber \\
&~& \left(T^\prime_{\rm a} \over 0.012\, {\rm s} \right)^{0.06\pm 0.59}.
\label{lz}
\end{eqnarray}
Note that the exponent of \tap\ is very close to zero and this implies that
{\it \tap\ does not reduce the scatter of the \epeiso\ correlation.}  Indeed,
the above result is almost identical to the \epeiso\ correlation defined by
the 23 GRBs reported in Tab. 1 (see also the last row of Tab. 2).

For consistency with the fits reported in the previous sections we also fitted
the \ep--\eiso--\tap\ correlation by weighting for the errors on the three
variables. In this case we find a different result:
\begin{eqnarray}
E_{\rm \gamma,iso, 52} \, &=& (13.8\pm 2.0)\, 
\left(E_{\rm p} \over 513\, {\rm keV} \right)^{2.18\pm 0.53} \, \times
\nonumber \\
&~&
\left(T^\prime_{\rm a} \over 0.0173\, {\rm s} \right)^{-0.67\pm 0.43}.
\label{lze}
\end{eqnarray}
with a reduced $\chi^2=2.28$ (20 dof). 
The reason for the difference with Eq. \ref{lz} lies in 
the errors on \tap, which are particularly large
(see Tab. 1).
To prove it, we have reduced the errors on \tap\ by a
factor 1.5 and recovered the result of the fit 
with a simple linear regression analysis (Eq. \ref{lz}).

In Fig. \ref{scatter} we compare the scatter of the data points around the
\epeiso\ correlation (defined by the 23 GRBs of Tab. 1) with the scatter of
the data points around the best fit plane of Eq. \ref{lze} (i.e.  obtained by
weigthing for the errors). We note that the scatter of the data points in the
\ep--\eiso--\tap\ plane is comparable to the scatter of the data points in the
\ep--\eiso\ plane. This suggests that, although in Eq. \ref{lze} (found by
weighting for the errors) the exponent of \tap\ is different from zero, the
scatter of the data points is not improved with respect to the \epeiso\ 
correlation.  We conclude that it is Eq. \ref{lz} which better represents the
best fit to the data.

\begin{figure}
\centerline{
\psfig{figure=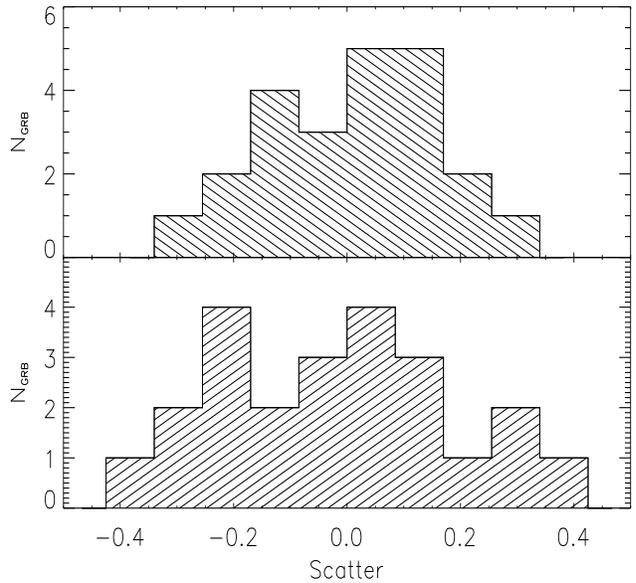,width=9cm,height=8cm}}
\caption{
  Scatter distribution of the 23 GRBs of Tab. 1 around the best fit \epeiso\ 
  correlation (top histogram) compared with the scatter of the same points
  around the best fit plane of the \ep--\eiso--\tap\ correlation defined by Eq.
  \ref{lze} (bottom histogram).  One can see that the scatter of the
  \ep--\eiso--\tap\ correlation is similar (if not larger) than the scatter of
  the \epeiso\ correlation.  }
\label{scatter}
\end{figure}

\begin{figure}
\centerline{
\psfig{figure=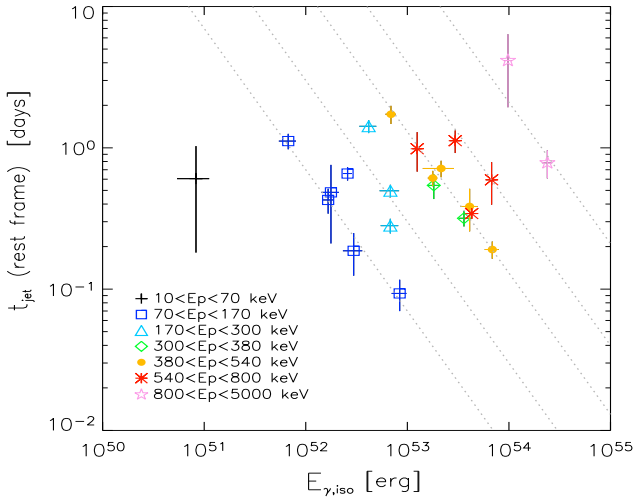,width=9cm,height=7cm}}
\caption{
  The time \tjet\ as a function of
  \eiso\ for the burst of the G07 sample. Bursts with different
  peak energy ranges (as labelled) are shown with different symbols. 
The dotted lines indicate \tjet$\propto E_{\rm \gamma, iso}^{-1}$,
with different normalizations, to illustrate that bursts with
the same \ep\ do follow this law.
}
\label{eiso_t}
\end{figure}
\begin{figure}
\centerline{
\psfig{figure=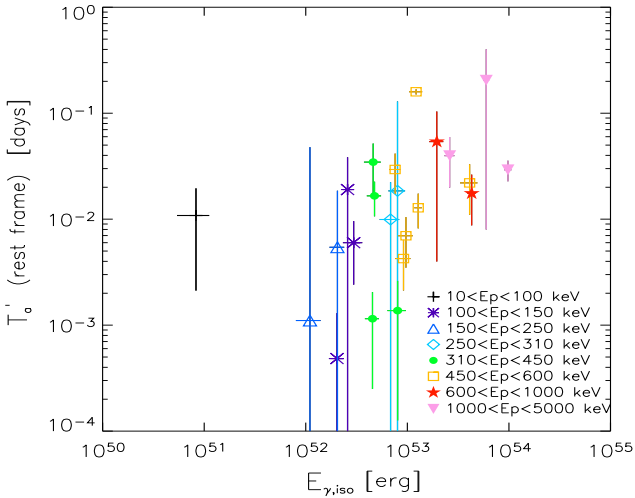,width=9cm,height=7cm}}
\caption{
  The time \tap\ vs \eiso\ for the burst of this paper (Tab. 1).
  As in Fig. \ref{eiso_t} we use different symbols for bursts 
  with \ep\ in different ranges.}
\label{eiso_ta}
\end{figure}

This would be sufficient to demonstrate that \ta\ does not help to define a
tighter correlation between the isotropic energy and \ep.  To further
demonstrate this point, we directly study the relation between \ta\ and \eiso,
comparing it with the corresponding relation between \tjet\ and \eiso.

Consider bursts with the same \ep. For these the existence of the empirical
Liang \& Zhang correlation implies that \eiso$\propto {t^{\prime}_{\rm
    jet}}^{-1}$.  This is proved by the data: in Fig. \ref{eiso_t} we show the
anticorrelation between \tjet\ (rest frame) and \eiso\ for the sample of 23
GRB by considering bursts with similar \ep\ (corresponding to the different
symbols in Fig. \ref{eiso_t}).  The \eiso$\propto {t^{\prime}_{\rm jet}}^{-1}$
relation converts in \eiso$\theta_{\rm jet}^2$=const (the ``Frail
correlation") when using the relation between \tjet\ and $\theta_{\rm j}$ (Eq.
1), for both the HM and the WM cases.

Consider now Fig. \ref{eiso_ta}, showing \tap\ versus \eiso: in this case
\tap\ does not anticorrelate with \eiso, for constant \ep.  The absence of the
anticorrelation translates in the impossibility to obtain a similar ``Frail
correlation" when using \ta\ as a jet angle indicator (i.e. inserting \ta\ 
instead of \tjet\ in Eq. 1).

\subsection{The origin of the \epega\ correlation}

Although \ta\ spans 3 orders of magnitudes, and although it is not a jet angle
indicator, the \epega\ correlation exists and has a scatter comparable (or
slightly larger) than the scatter of the \epeiso\ correlation.  What is, then,
its origin?

In Eq. \ref{ang} the jet angle depends on both the break time
(either \ta\ or \tjet) but also on $E_{\gamma, \rm iso}/\eta_\gamma$.  This
term is the kinetic energy of the GRB outflow left--over after the prompt
radiative phase.  Since, usually, one assumes a constant value
$\eta_{\gamma}=0.2$, the collimation correction depends on \eiso.

In this section we explore the role of \eiso\ and $t_{\rm break}$ (either
identified as \ta\ or \tjet) in Eq. \ref{ang} in shaping the \epega\ and the
\epeg\ correlation and their scatter.

From Eq. \ref{ang} considering only the dependencies from \eiso\
and $t_{\rm break}$ we find (under the assumption of small angles):
\begin{eqnarray}
\rm WM:   \eg & \propto & \eiso~\theta^2\propto \eiso~\frac{t_{\rm
    break}^{1/2}}{E^{1/2}_{\rm \gamma,iso}}\propto E^{1/2}_{\rm
    \gamma,iso}t_{\rm break}^{1/2} \nonumber \\
\rm HM:   \eg & \propto &\eiso~\theta^2\propto \eiso~\frac{t_{\rm
    break}^{3/4}}{E^{1/4}_{\rm \gamma,iso}}\propto E^{3/4}_{\rm
    \gamma,iso}t_{\rm break}^{3/4}
\label{te}
\end{eqnarray}

Consider those GRBs that have the same $E_{\rm p}$,
but different \eiso.
Even if they have the same $t_{\rm break}$,
the use of Eq. \ref{te} implies a clustering of the
corresponding collimation corrected $E_\gamma$:
GRB with larger \eiso\  would have smaller
$\theta$, and therefore have a larger correction
for their collimation (and vice-versa).

We calculate $\theta$ and correct for the collimation factor {\it assuming
  that the break time (either \ta\ or \tjet) is constant and equal for all
  bursts.}  The use of Eq. \ref{te} with a fixed $t_{\rm break}$ gives
$E_{\gamma}\propto E_{\gamma,\rm iso}^{1/2}$ in the WM case, and
$E_{\gamma}\propto E_{\gamma,\rm iso}^{3/4}$ in the HM case.  We plot \ep\ 
versus this estimate of $E_{\gamma}$ in Fig. \ref{test} (for both the WM and
HM cases).  The empty star symbols define a correlation which is less
scattered than the \epeiso\ correlation (empty circles) in the HM and WM case
(left and right panels in Fig. \ref{test}).  Moreover, the scatter reduction
is somewhat more significant in the WM case with respect to the HM case.

Note that this result has no physical meaning and it is only due to the fact
that we plot the points in the \ep-$E^{1/2}_{\rm \gamma,iso}$ plane for the WM
case (or \ep--$E^{3/4}_{\rm \gamma, iso}$ for the HM case) instead of in the
\ep--\eiso\ plane.  Up to this point the reduction of the scatter of the
\epeiso\ correlation, and the difference in the slope with respect to the Amati
relation, is due to the dependence of Eq. \ref{te} (used to derive the jet
opening angle) on \eiso.

We can now see if the use of the real break time (either \ta\ or \tjet\ --
upper panels and lower panels in Fig.  \ref{test}, respectively) improves or
not the found correlations.  The results in the case of \ta\ are shown in the
upper panels of Fig. \ref{test} for the HM and WM case.  Although a
correlation between \ep\ and $E_{\rm \gamma,a}$ appears (filled circles), we
find that the effect of \ta\ is to {\it increase} the scatter of the data
points with respect to the same correlation defined with \ta\ constant.  Note
that the scatter of the found \epega\ correlation is not less than that of the
\epeiso\ correlation (both in the HM and WM case).  This demonstrates that the
variable \ta, if treated as a jet break time to estimate the angle and derive
the collimation corrected energy, leaves unaltered (or even increases) the
scatter of the \epeiso\ correlation.  This effect is clearly shown in Fig.
\ref{test} by bursts with typical rest frame peak energy in the range
100--1000 keV which are the majority in our sample.

\begin{figure*}
\vskip 0.5 true cm
\centerline{
\psfig{figure=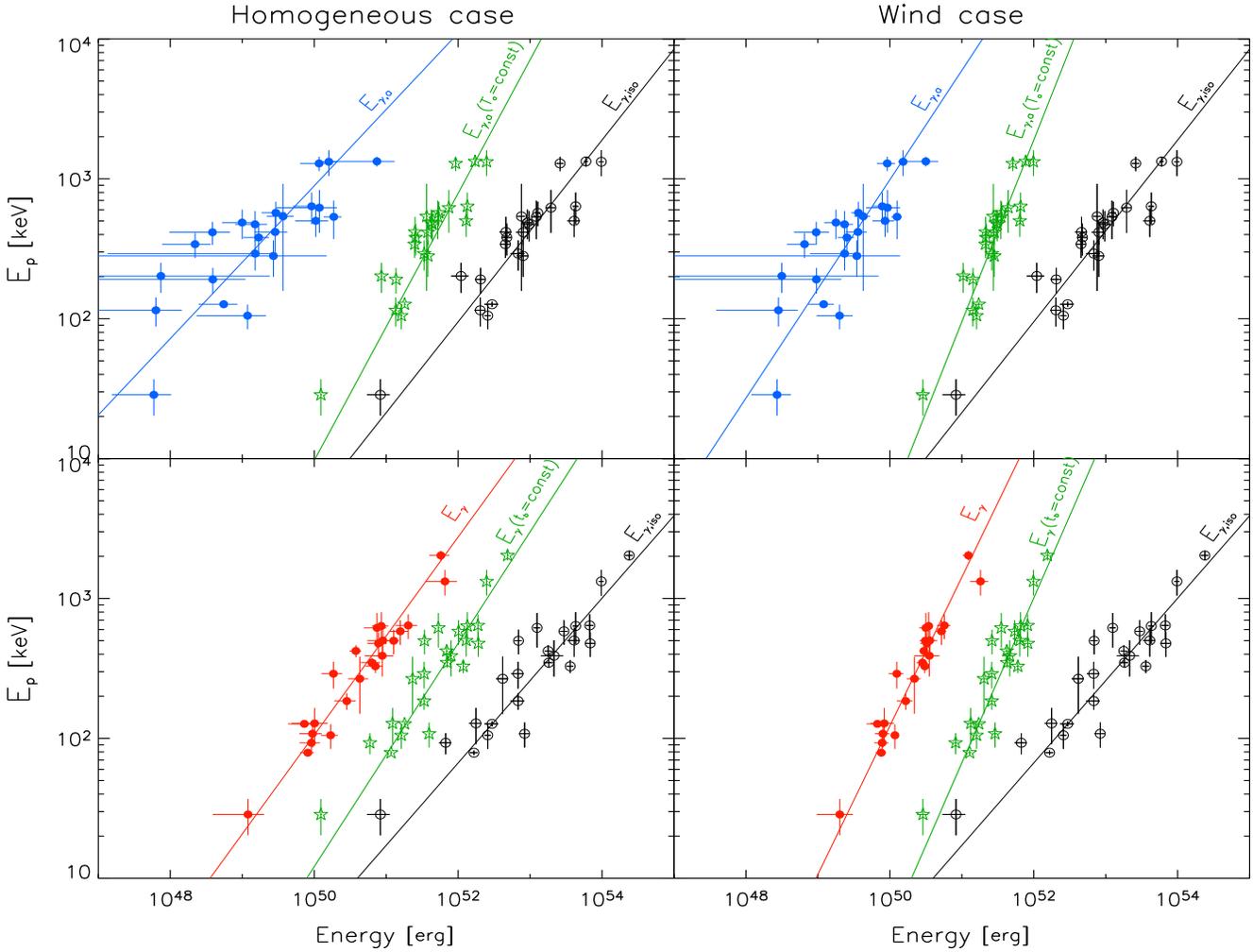,width=18cm,height=13cm}}
\caption{
  Comparison of the scatter of the \epega\ and \epeg\ correlations in the
  homogeneous (left) and wind (right) medium cases.  Top panels: from right to
  left we report the \epeiso\ correlation (empty circles), the correlation
  defined by assuming a constant value of \ta\ (for all the bursts) in
  computing the collimation correction (empty stars) and the correlation
  defined by using the measured \ta\ (filled circles; values of \ta\ reported
  in Tab. 1).  Bottom panels: same as above but considering \tjet\ instead of
  \ta. }
\label{test}
\end{figure*}

In the bottom panels of Fig. \ref{test} we show the results obtained with
\tjet.  As discussed in Sec. 2, the sample of bursts used to define the \epeg\ 
correlation only marginally overlaps to that used to define the \epega\ 
correlation: this is the reason why also the \epeiso\ correlation in the lower
and upper panels of Fig. \ref{test} are different.  If first we set
\tjet=constant and equal for all bursts we have, as before, a reduction (empty
stars) of the scatter of the \epeiso\ correlation (empty circles).  But now,
if we assign to each GRB in our sample its \tjet\ found from the optical light
curve, we see that the scatter of the data points is {\it further reduced}
(note in particular this effect in peak energy range 500--1000 keV).

From these results we conclude that:
\begin{itemize}
\item the dependence of the jet angle on \eiso\ reduces the scatter of
  the \epeiso\ correlation;
\item the use of the time \ta\, which marks the end of the plateau phase in
  the X--ray light curves, increases the scatter of the correlation found
  assuming a constant \ta. This effect, counterbalanced by the previous
    one, makes the scatter of the \epega\ correlation similar to the scatter
  of the \epeiso\ correlation;
\item the (albeit weak) positive correlation between \ta\ and \eiso\ (see Fig.
  \ref{eiso_ta}), makes the slope of the \epega\ correlation to differ
  somewhat from the slope of the correlation found with \ta\ constant;
\item the small dispersion of the Ghirlanda relation is due not only to the
  presence of \eiso\ in the equation for the angle, but also to the jet break
  time \tjet.
\end{itemize}

\begin{figure}
\vskip -0.5 true cm
\hskip -0.3 true cm
\centerline{
\psfig{figure=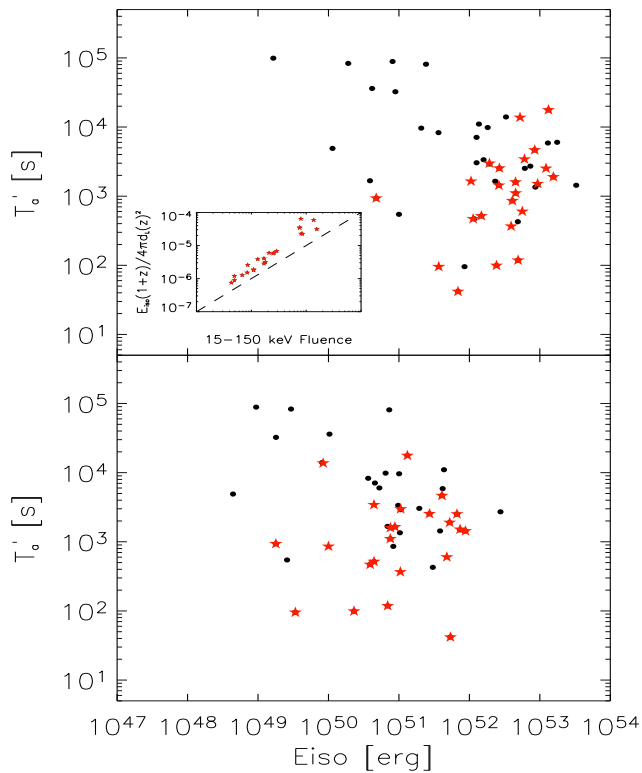,width=9.5cm,height=11cm} }
\caption{
Top panel: the time  \tap\ as a function of the isotropic energy of the
prompt emission \eiso\ calculated in the 15--150 keV
energy range, without including any bolometric and $k$--correction
for all bursts with measured redshift (circles) and for
those which also have a  measured \ep\ (stars).
The insert shows the bolometric fluence vs the 15--150 keV fluence,
to show that the two are proportional. Bursts in the insert are
the ones in Tab. 1.
Bottom panel: the time \tap\ as a function of the energy emitted
in the plateau phase, in the 0.3--10 keV energy range.
The fluences of the plateau phase have been taken from W07.
Symbols as above.
}
\label{fluence}
\end{figure}

\section{Summary and discussion}

We have analyzed in detail the correlation found by W07
between $E_{\rm p}$ and the collimation corrected
energy $E_{\rm \gamma, a}$, found using \ta\ as a jet break.
Our sample (of 23 objects) is larger than what used by W07,
because we could add some GRB occurred after August 2006,
and some other bursts for which we could find the value of \ep.
We find that \ep\ and \ega\ correlate, by we also find that
this correlation is weaker, and has a different slope,
than the \epeg\ one.

Investigating further, we have also shown that the
\epega\ correlation is entirely a consequence of
the existence of the \epeiso\ (Amati) correlation,
coupled with the fact that the opening angle
$\theta_{\rm j}$ depends on \eiso.
This is why one finds a correlation which is 
narrower than the Amati one even using the same
\ta\ for all bursts: {\it in this case the found
correlation is even better than the one obtained
introducing the real \ta.}
This demonstrates that \ta\ does not play any
role in the \epega\ correlation: it is even
worsening it.

We have then made the very same test to the \epeg\ (Ghirlanda)
correlation, and found that in this case the use of the
``real" \tjet\ improves the correlation, decreases the 
scatter and slightly changes the slope of the correlation,
with respect to the one obtained using a fixed \tjet.

All these results are fully confirmed by analyzing the
\ep--\eiso--\tap\ correlation, and comparing it to the 
Liang \& Zhang correlation (which uses \tjet\ instead of \ta).
Again, all evidence is towards \tap\ playing no role:
the found \ep--\eiso--\tap\ correlation has a similar
(if not larger) scatter than the \epeiso\ correlation.

We conclude that the time \ta\ at which the plateau
of the X--ray light curve ends is not an important
parameter for defining the spectral--energy correlations.
As a consequence, it is not a jet angle indicator.
Instead, we confirm that 
\tjet\ is likely to measure the jet opening angle,
since it passed our simple test to see if it
helps or not to strengthen the correlation between
\ep\ and the collimation corrected energy derived
using a fixed value of \tjet\ for all bursts.

We note a weak correlation between \tap\ and \eiso\ when considering the 23
GRBs in our sample (see Fig. \ref{eiso_ta}, probability of chance correlation
equal to $4\times 10^{-3}$, according to the Spearman rank correlation), which
however disappears when considering the entire sample of GRBs of known
redshifts and \tap\ (and relaxing the requirement to have also \ep\ known, see
Fig. \ref{fluence}).  Therefore we can also conclude that \tap\ is not related
to \eiso, differently from \tjet\ which is instead anti--correlated with
\eiso\ when considering GRBs of similar \ep.  Note that, for a distribution of
\tap\ not correlated (or anti--correlated) with \eiso, the resulting \epega\ 
correlation should become parallel to the one obtained fixing \ta\ to a
constant value.  We therefore predict that, for larger samples of GRBs with
known redshift, \ta\ and \ep, the \epega\ correlation will have a slope
$\approx 0.75$ ($\approx 1$) in the HM (WM) case.

Finally, there is no correlation between \tap\ and the
isotropic energy (in the observed 0.3--10 keV energy range)  
emitted during the plateau phase (see Fig. \ref{fluence}).

Unfortunately, our finding that the \epega\ correlation is
only a by--product of the \epeiso\ (Amati) correlation
is not very helpful in shedding new light on the problem
of explaining why the light curve of most GRBs is characterized
by the steep--flat--steep behaviour, and also why this is 
sometimes different by the simultaneous behavior observed 
in the optical band.

Recently, we (Ghisellini et al. 2007) have proposed that the
X--ray light curve is produced by internal dissipation in late
shells, where ``late" means that the central engine
continues to create relativistic shells even days after the trigger.
This implies that the central engine is long lived, and that there
is a sharp transition between the early phase  (determining the 
duration of the early prompt, i.e. the time $T_{90}$), and the
late phase, emitting much smaller luminosity, but for a longer time.
The early phase should be characterized by a large bulk Lorentz
factor $\Gamma$ of the created shells, changing erratically, while 
the late phase is instead characterized by shells created with 
$\Gamma$ decreasing monotonically.
This implies that when $\Gamma>1/\theta_{\rm j}$, we see only a fraction 
of the emitting surface, but when $\Gamma\le 1/\theta_{\rm j}$ we see
the entire emitting surface.
This means that there should be a break in the light curve 
when $\Gamma=1/\theta_{\rm j}$.
We associate this time to \ta.

According to this idea, the two times \ta\ and \tjet\ have an obvious link:
both are determined by the jet opening angle $\theta_{\rm j}$.  Then, why we
do not observe a relation between these two times?  We propose the following
answer: the time \tjet\ is determined when the fireball made by the merged
shells of the early phase, carrying a kinetic energy $E_{\rm k}$, is
decelerated by the external medium.  The dynamics of the process is robust,
depending only on the value of the external medium density (and its profile
with distance), on $E_{\rm k}$, and finally on the conservation of energy and
momentum.  We have the same $\Gamma(t)$ law for different GRBs, if the
external medium has the same profile.

On the contrary, the law characterizing how $\Gamma$ changes as a function 
of time during the late prompt phase can be different from burst to burst.
As far as we know, there might not be a robust analog of the conservation
of energy and momentum dictating the $\Gamma(t)$ behaviour in this case.
Indeed, we can have information on this law from the observed slopes
of the light curve (see Ghisellini et al. 2007), and under 
simplifying hypotheses we can already derive that $\Gamma(t)$ is different
from one bursts to another.
As a consequence, there is no general formula,
equal to all bursts, in which \ta\ can be inserted 
to derive the jet opening angle.

\section*{Acknowledgements}
We thank F. Tavecchio for useful discussions and the anonymous referee for
his/her suggestions. PRIN--INAF is thanked for a 2005 funding grant.


\begin{thebibliography}{}

\bibitem[]{} Amati, L., Frontera, F., Tavani, M., et al. 2002, A\&A, 390, 81
\bibitem[]{} Amati, L., 2006, MNRAS, 372, 233

\bibitem[]{} Band D. L., et al., 1993, ApJ, 413, 281
\bibitem[]{} Barbier, L., Barthelmy, S., Cummings, J., et al. 2006, GCN, 4518
\bibitem[]{} Bellm, E.,  Bandstra, M., Boggs, S., Wigger, C., Hajdas, W., 
             Smith, D.M., \& Hurley, K. 2006, GCN, 5838
\bibitem[]{} Blustin, A.J., Band, D., Barthelmy, S., et al. 2006, ApJ, 637, 901

\bibitem[]{} Burrows et al., 2005, Sci, 309, 1833
\bibitem[]{} Butler, N.R. \& Kocevski, D., 2007, subm to ApJ (astro--ph/0612564)
\bibitem[]{} Cenko, S.B., Kasliwal, M., Harrison, F.A., et al. 2006, ApJ, 652, 490
\bibitem[]{} Crew, G., Ricker, G., Atteia, J-L., et al. 2005, GCN, 4021

\bibitem[]{} Curran, P., Kann, D.A., Ferrero, P., Rol, E. \& Wijers R.A.M.J.,
             2006, Il Nuovo Cimento C., in press (astro--ph/0611189)
\bibitem[]{} Firmani et al., 2006, MNRAS, 370, 185

\bibitem[]{} Frail D. et al., 2001, ApJ, 562, L55

\bibitem[]{} Gehrels, N., Chincarini, G., Giommi P., et al., 2004, ApJ, 611, 1005

\bibitem[]{} Ghirlanda, G., Ghisellini, G. \& Lazzati, D., 2004,  ApJ, 616, 331 (GGL04)
\bibitem[]{} Ghirlanda, G., Nava, L., Ghisellini, G. \& Firmani, C., 2007,
             A\&A subm (G07)
\bibitem[]{} Ghisellini, G., Ghirlanda, G., Nava, L. \& Firmani, C., 2007, 
             ApJ Lett. subm. (astro--ph/0701430)
\bibitem[]{} Golenetskii, S., Aptekar, R., Mazets, E., Pal'shin, V., Frederiks, D., \& Cline, T. 2005a, GCN, 3179
\bibitem[]{} Golenetskii, S., Aptekar, R., Mazets, E., Pal'shin, V., Frederiks, D., \& Cline, T. 2005b, GCN, 3518
\bibitem[]{} Golenetskii, S., Aptekar, R., Mazets, E., Pal'shin, V., Frederiks, D., \& Cline, T. 2005c, GCN, 4238
\bibitem[]{} Golenetskii, S., Aptekar, R., Mazets, E., Pal'shin, V., Frederiks, D., Ulanov, M., \& Cline, T. 2006a, GCN, 4989
\bibitem[]{} Golenetskii, S., Aptekar, R., Mazets, E., Pal'shin, V., Frederiks, D., \& Cline, T. 2006b, GCN, 5837

\bibitem[]{} Liang, E. \& Zhang, B., 2005, ApJ, 633, L611  

\bibitem[]{} Kumar, P. \& Panaitescu, A., 2000, ApJ, 647, 1213
\bibitem[]{} Nava, L., Ghisellini, G., Ghirlanda, G., Tavecchio, F. \& Firmani, C.,
             2006, A\&A, 450, 471 (N06)

\bibitem[]{} Nousek J., et al., 2006, ApJ, 642, 389
\bibitem[]{} O'Brien et al., 2006,, ApJ, 647, 1213

\bibitem[]{} Palmer, D., Barbier, L., Barthelmy, S., et al. 2006, GCN, 4697
\bibitem[]{} Perri, M., Giommi, P., Capalbi, M., et al. 2005, A\&A, 442, L1
\bibitem[]{} Press, W.H. et al. 1999, {\it Numerical Recipes in C}, 
             Cambridge University Press, 661

\bibitem[]{} Rees, M.J. \& Meszaros, P., 1992, MNRAS 258, L41
\bibitem[]{} Rhoads J., 1997, ApJ, 487, L1
\bibitem[]{} Romano, P., Campana, S., Chincarini, G., et al. 2006, A\&A, 456, 917
\bibitem[]{} Sato, G., Yamazaki, R., Ioka, K., et al. 2006, astro-ph/0611148, accepted for publication in ApJ
\bibitem[]{} Schaefer, B.E. 2007, ApJ, in press (astro--ph/0612285)
\bibitem[]{} Stamatikos, M., Barbier, L., Barthelmy, S., et al. 2006a, GCN, 5289
\bibitem[]{} Stamatikos, M., Barbier, L., Barthelmy, S., et al. 2006b, GCN, 5639
\bibitem[]{} Willingale, R., et al., 2007, subm to ApJ (W07) (astro--ph/0612031)
 
\end{thebibliography}
\end{document}